\documentclass{aa}
\usepackage{amsmath}
\usepackage{graphicx}
\usepackage{natbib}
\usepackage{textcomp}
\usepackage{color}
\usepackage[tableposition=top]{caption}
\usepackage[varg]{txfonts}
\begin{document}
\title{The nature of separator current layers in MHS equilibria}
\titlerunning{The nature of separator current layers in MHS equilibria}
\subtitle{I. Current parallel to the separator}
\author{J. E. H. Stevenson, C. E. Parnell, E. R. Priest and A. L. Haynes }
\authorrunning{J. E. H. Stevenson}
\institute{School of Mathematics and Statistics, University of St Andrews, North Haugh, St Andrews, Fife, KY16 9SS, Scotland\\ \email{jm686@st-andrews.ac.uk}}
\date{}
\abstract{Separators, which are in many ways the three-dimensional equivalent to two-dimensional nulls, are important sites for magnetic reconnection. Magnetic reconnection occurs in strong current layers which have very short length scales.}{The aim of this work is to explore the nature of current layers around separators. A separator is a special field line which lies along the intersection of two separatrix surfaces and forms the boundary between four topologically distinct flux domains. In particular, here the current layer about a separator that joins two 3D nulls and lies along the intersection of their separatrix surfaces is investigated.}{A magnetic configuration containing a single separator embedded in a uniform plasma with a uniform electric current parallel to the separator is considered. This initial magnetic setup, which is not in equilibrium, relaxes in a non-resistive manner to form an equilibrium. The relaxation is achieved using the 3D MHD code, Lare3d, with resistivity set to zero. A series of experiments with varying initial current are run to investigate the characteristics of the resulting current layers present in the final (quasi-) equilibrium states.}{In each experiment, the separator collapses and a current layer forms along it. The dimensions and strength of the current layer increase with initial current. It is found that separator current layers formed from current parallel to the separator are twisted. Also the collapse of the separator is a process that evolves like an infinite-time singularity where the length, width and peak current in the layer grow slowly whilst the depth of the current layer decreases.}{}
\maketitle
\section{Introduction}\label{sec:introduction}
The importance of the fundamental energy release mechanism called magnetic reconnection is made apparent by the key role it plays in many plasma processes on the Sun and other stars (e.g., coronal mass ejections, coronal heating, solar and stellar flares) and in the magnetosphere (e.g., powering flux transfer events and substorms) \citep[e.g.,][]{Biskamp2000, PriestForbes}. Magnetic reconnection permits the restructuring of the magnetic field enabling changes in magnetic topology or quasi-topology to occur. Reconnection converts magnetic energy to thermal energy, kinetic energy (bulk plasma motions) and fast particle energy. The partitioning of magnetic energy into these three forms depends on the nature of the reconnection itself and the properties of the surrounding plasma. 

Reconnection in two dimensions (2D), which was first proposed as a mechanism for flares in the 1940's \citep{Giovanelli46,Hoyle49}, has been studied in detail since the 1950's \citep[e.g.,][]{Parker57,Sweet58,Biskamp1982,Priest1986,Biskamp2000, PriestForbes}. More recently, three dimensional (3D) reconnection has been explored and has proved to be much more complex than 2D reconnection due to the multitude of possible reconnection sites, the different types of reconnection (null-point, separator, quasi-separator) and the increased intricacy of the 3D magnetic skeleton.

The lowest energy state of any magnetic field, ${\bf B}$, in a closed volume with the normal component imposed on the boundary is its potential (current-free) field, in which $\nabla\times{\bf B}={\bf 0}$. Electric currents, ${\bf j}$, will always be present in any magnetic field that is not at its lowest energy.  Furthermore, the magnetic Reynolds number, $R_m = vL/\eta$ where $v$ and $L$ are typical velocity and length scales in the system and $\eta$ is the magnetic diffusivity, is normally much larger than unity and represents the ratio of the advection and diffusion terms in the induction equation. Thus reconnection, which requires the magnetic field to be able to diffuse and, hence, $R_m \ll 1$, requires short length scales. Thus current concentrations, in which there are steep gradients in the magnetic field over short length scales, are sites in the solar atmosphere (and throughout the Universe) where reconnection is likely to occur. In this paper, we concern ourselves with the properties of 3D current concentrations formed at separators through the non-resistive relaxation of the magnetic field, rather than the nature of the reconnection that occurs within them. 

In 2D, current layers are known to form following the collapse of 2D null points. This has been studied in detail both in the zero-beta approximation \citep[e.g.,][]{Green65,Syrovatskii71,Somov76,Craig94,Bungey95} and in the non-zero beta approximation \citep[e.g.,][]{Rastatter94, Craig05, Pontin05, FF11}. The key features associated with these types of current layers are (i) a collapse of the separatrices through the null forming cusp regions, (ii) enhanced current about the null point and along the separatrices extending beyond the ends of the main current layer and (iii) higher density plasma within the cusp regions than outwith them. \cite{Klapper98} showed analytically that in 2D, in the zero-beta approximation, a current layer at a null point will never reach equilibrium, since the collapse time of the null is infinite. This fact is still true when the plasma beta is non-zero \citep{Craig05, Pontin05, FF11}. \cite{FF11}, who studied the magnetohydrodynamic (MHD) collapse of 2D nulls in the absence of resistivity, showed that a state may be reached in which the magnetic field and plasma are close to equilibrium everywhere save within the highly localised current accumulation itself.

In 3D, current layers are also likely to form. In general, however, most models of 3D reconnection have considered driven reconnection. These experiments typically start from a potential field and initiate reconnection by driving at the boundaries at either a fast or slow rate. Aspects of the nature of the reconnection in these models are dependent then on not only the initial magnetic field, but also are heavily dependent on the rate and nature of the boundary drivers. 3D reconnection can occur at null points, as does 2D reconnection, and this has been studied under different driven regimes, \citep[e.g.,][]{Craig95, Craig96, Priest96, pontin04a, Pontin05, pontin05c, pontin05b, Pontin07, Masson09, priest09a, pontin11b, Masson12, pontin13a}. 3D reconnection can also occur in the absence of null points \citep{Schindler88,Hesse88}, for instance, at separators \citep[e.g.,][]{Galsgaard96,Priest96,Longcope96, Longcope01, Haynes07, Parnell10a, Parnell10b, Wilmot11} or at quasi-separatrix layers \citep[e.g.,][]{Demoulin95,Demoulin96,Demoulin97,Aulanier06,Wilmot09}.
 
It is, however, generally believed that in plasma systems, such as the solar atmosphere or Earth's magnetosphere, the stressing of magnetic structures due to the slow driving of magnetic field lines leads to a build up of free magnetic energy (the excess energy above potential) associated with electric currents. If topologically or geometrically complex 3D magnetic fields are stressed then the equilibria that are formed will have current layers located, for example, where the field-line mapping is discontinuous or has strong gradients. 

Eventually, the length-scales within these current layers become sufficiently short that the magnetic Reynolds number is less than or equal to one, allowing reconnection to occur (initiated, for instance, by micro-instabilities). However, practically all models of reconnection (whether simulations of solar flares or experiments studying aspects of the fundamental physics of reconnection), such as many of those mentioned previously, start from an initial potential magnetic field, i.e. one with no free energy.

Our ultimate aim is to study spontaneous reconnection, as opposed to driven reconnection, in systems which have excess magnetic energy stored in current layers that are in equilibrium with their surroundings. In order to study this type of reconnection, the initial magnetohydrostatic (MHS) equilibria with current layers needs to be created. Non-zero beta MHS equilibria involving a current layer situated at a 2D null have been studied by \cite{Craig05, Pontin05, FF11} with the resulting spontaneous reconnection studied by \cite{FFP12, FFP12b}. Current layers in 3D systems have also been considered. These include current layers generated by (i) the shearing of uniform magnetic fields \citep{Longbottom98,Bowness},  (ii) the tangling of multiple flux tubes \citep{wilmot09a,Wilmot09}, (iii) at 3D null points \citep{Pontin05,FF12,FF13},  (iv) current layer formation due to ideal MHD instabilities \citep{Browning08} and (v) at quasi-separatrix layers \citep{GTN03, TGN03, aulanier05a,wilmot09b}. However, MHS equilibria with current layers situated on magnetic separators have never been studied before. So, in this paper, we study these equilibria. In a follow up paper, we will look at the nature of the spontaneous reconnection that occurs in these single-separator current-layer systems.   

Although 3D null points are in some ways the natural equivalent to 2D nulls, in other ways 3D magnetic separators are their equivalent. Generic separators are special field lines that are the intersection of two separatrix surfaces\footnote{Separators may also be formed by the intersection of the separatrix surface from one null with the spine of another, or by the intersection of the spines from two separate nulls. These two types of separators are non-generic as they are topologically unstable since a slight perturbation could lead to the intersection, and hence the separator, no longer existing. Therefore, in this paper, we only consider generic separators which are formed by the intersection of two separatrix surfaces.}, such as bald-patch separatrices or fan surfaces. In the latter case, the separators run from one 3D null point to another. Thus, like 2D nulls, separators lie at the boundary between four topologically distinct flux domains \citep{Priest96,Longcope98, Haynes07}. Also, perpendicular cuts across a separator reveal that the projection of the magnetic field lines in these planes can be hyperbolic or elliptic which is analogous to the magnetic field structure about a 2D X-point or O-point, respectively \citep{Parnell10a}. 

When separator reconnection occurs, flux from two oppositely situated flux domains moves into the other two domains, which is akin to what is observed at 2D null-point reconnection. However, in 3D this reconnection does not involve field lines that reconnect one pair at a time to create a new pair of field lines. Instead, whole surfaces counter-rotate about the separator reconnecting a continuum of field lines (but containing a finite amount of flux). Furthermore, in the same way that 2D reconnection leads to a discontinuous mapping of the field lines, so also separator reconnection is associated with a discontinuous field line mapping. 
Numerical experiments \citep{Haynes07,Parnell10b,Parnell2011} and analytical work \citep{Wilmot11} reveal that separator reconnection is quite different from 3D null-point reconnection. The field lines reconnect at some location along the separator, where the parallel electric field (parallel current) is enhanced, away from the null points.

Magnetic separators have been recognised as important locations of 3D reconnection for many years, as current builds up easily along them due to their special situation at the boundary between topologically distinct domains \citep[e.g.,][]{Sonnerup79,LauFinn,Priest96,Haynes07,Parnell2011}. However, the nature of current accumulations in the vicinity of separators has not yet been properly investigated. 

Here, we study the MHS equilibria that are created through the non-resistive MHD relaxation of a non-potential magnetic field containing two 3D null points connected by a separator. As we will show, these equilibria involve current layers embedded in potential magnetic fields exactly as is found in the collapse of both 2D \citep[e.g.,][]{Pontin05,FF11} and 3D nulls \citep[e.g.,][]{FF12,FF13}. 
 The numerical model, Lare3d, used to perform the relaxation, is detailed in Sect.~\ref{sec:lare}. The setup and properties of the initial magnetic field and plasma are described in Sect.~\ref{sec:initial}. A series of experiments have been performed which all start from the same initial setup, save for the initial current which differs. The final equilibrium configurations of all the experiments have current accumulations with the same basic nature and characteristics. Thus, Sect.~\ref{sec:results} highlights the nature of the final (quasi-) equilibrium state achieved after the relaxation in one particular example experiment. Sect.~\ref{sec:vary} then considers the effects of varying the initial current and compares the characteristics of the current layers formed in each relaxation. We conclude with a discussion of our findings in Sect.~\ref{sec:conc} which is followed by an appendix (Appendix~\ref{sec:app}) where further details of the initial non-potential magnetic field used for the relaxation experiments can be found.

\section{Numerical model}\label{sec:lare}
We are interested in determining the current accumulations that occur due to the non-resistive MHD relaxation of an initial non-potential magnetic field involving a separator. Our focus is not on the process of the MHD relaxation (although this is discussed briefly, in Sect.~\ref{sec:energy}), but on the characteristics of the final MHS equilibria. 
The initial system we start from is not in force balance so, as soon as it starts to relax, waves are generated. These waves are damped due to the presence of viscosity and so, to generate our magnetic equilibria, we used a 3D non-resistive MHD code, namely, Lare3d \citep{Arber01}. 

Lare3d is a staggered Lagrangian re-map code, in which the scalar quantities ($\rho$ - density, $\epsilon$ - internal energy per unit mass and $p$ - pressure) are defined at the cell centres and the magnetic field components, ${\bf {B}}$, are defined on the cell faces to help maintain $\nabla\cdot{\bf {B}} = 0$. This is done using the Evans and Hawley constrained transport method for the magnetic flux \citep{Evans88}. Also, the velocity components, ${\bf {v}}$, are staggered with respect to the pressure and magnetic field to prevent checkerboard instabilities and so are placed at the cell vertices \citep{Arber01}. 
Lare3d works in two steps: (i) a LAgrangian step, where the MHD equations are solved in a frame that moves with the fluid; (ii) a REmap step, which is purely a geometrical mapping of the Lagrangian grid back onto the original Eulerian grid. Lare3d solves the normalised MHD equations and employs the following normalised quantities (using subscript $n$ to denote the normalising factors and hats to represent the dimensionless variables used by the code)
\begin{equation}
{\bf {x}} = L_{n}\hat{\bf {x}}, \qquad {\bf {B}} = B_{n}\hat{\bf {B}} \qquad\text{and}\qquad \rho = \rho_{n}\hat{\rho},
\end{equation}
where ${\bf {x}}$ is the length. These three normalising factors then define the following normalising constants for the velocity, pressure, current and internal energy per unit mass, respectively,
\begin{equation}
v_{n} = \frac{B_{n}}{\sqrt{\mu_{0}\rho_{n}}}, \hspace{0.2cm}  p_{n} = \frac{B_{n}^2}{\mu_{0}}, \hspace{0.2cm} j_{n} = \frac{B_{n}}{\mu_{0}L_{n}} \hspace{0.2cm}\text{and}\hspace{0.2cm}  \epsilon_{n} = v_{n}^2 = \frac{B_{n}^2}{\mu_{0}\rho_{n}},
\end{equation}
where $\mu_{0}$ is the magnetic permeability which is equal to 1 in dimensionless units. From these equations the plasma beta can be written as
\begin{equation}
\beta = \frac{2\hat{p}}{\hat{B}^2}.
\end{equation}
Therefore, in the absence of gravity and resistive effects, the standard normalised MHD equations used in Lare3d (with the hats dropped from the normalised quantities for ease of reading) are
\begin{equation}
\frac{\rm{D}\rho}{\rm{D}t} = -\rho \nabla \cdot {\bf{v}},
\end{equation}
\begin{equation}
\frac{\rm{D}{\bf{v}}}{\rm{D}t} = \frac{1}{\rho}(\nabla \times {\bf{B}}) \times {\bf{B}} - \frac{1}{\rho} \nabla p + \frac{1}{\rho}{\bf{F}}_{\nu},
\end{equation}
\begin{equation}
\frac{\rm{D}{\bf{B}}}{\rm{D}t} = ({\bf{B}} \cdot \nabla){\bf{v}} - {\bf{B}}(\nabla \cdot {\bf{v}}),
\end{equation}
\begin{equation}
\frac{\rm{D} \epsilon}{\rm{D}t} = -\frac{p}{\rho}\nabla \cdot {\bf{v}} + \frac{1}{\rho}H_{\nu},
\label{eq:inte}
\end{equation}
where ${\bf{F}}_{\nu}=\rho \nu(\nabla^{2}{\bf{v}}+\tfrac{1}{3}\nabla(\nabla \cdot {\bf{v}}))$ is the viscous force (where $\nu$ is the coefficient of kinematic viscosity) and $H_{\nu}=\rho \nu(\tfrac{1}{2}e_{ij}e_{ij} - \tfrac{2}{3}(\nabla \cdot {\textbf{v}})^2)$ is the viscous heating term (where $e_{ij} = (\partial v_i/\partial x_j)+(\partial v_j/\partial x_i)$ is the rate of strain tensor). To provide closure to these equations we require an equation of state: $\epsilon = p/(\rho(\gamma - 1))$, where $\gamma=5/3$ is the ratio of specific heats. 

The MHD equations that are usually solved by Lare3d contain resistive terms. However, to remove resistive effects from our MHD relaxation experiments, we remove these terms by setting the resistivity $\eta$ in the code to zero. Of course, though, all numerical codes suffer from numerical resistivity. In our code, we estimate the background numerical resistivity to be $\approx$ 0.0002. This is very small and we find that the numerical diffusion in our runs is negligible (see Sect.~\ref{sec:results} for a detailed discussion on this issue). The viscosity in the code is set to $\nu = 0.01$. 

The dimensions of the box are $-L_0$ to $L_0$ in the $x$ and $y$ directions, $-L_0$ to $L_0+L$ in the $z$ direction and the resolution of the grid is $512^3$. We have chosen this grid resolution because it is large enough to allow the experiments to evolve for a sufficient length of time for a current layer to form. We were, however, restricted in how large a grid resolution was feasible by e.g., memory and running time of the experiment.

The boundary conditions are chosen to prevent energy leaving or entering the domain. Thus the magnetic field is line tied at the boundaries and the scalar quantities (internal energy per unit mass and density) have a maximum or minimum on the boundary: i.e., the derivatives across the boundary of all three components of ${\bf B},\;\rho$ and $\epsilon$ are set to zero. The velocity (all components) is set to ${\textbf{0}}$ on all the boundaries. 
\section{Initial setup}\label{sec:initial}
\subsection{Magnetic field}
In our experiments, we study the nature of the current layer created in a system involving a single separator, which naturally also has two null points with associated spines and fans. We do not wish to influence where the current layer(s) form, therefore, we start initially with a uniform current throughout our domain so that during the relaxation the current has the freedom to choose where it collects: at the nulls, spines, fans or the separator.
  
The initial magnetic field we use contains just two null points connected by a single separator. It can be written analytically as follows:
\begin{eqnarray}
&& B_{x}=\tfrac{B_0}{L_{0}}(x+cxz+byz-\tfrac{1}{2} j_{sep}y), \nonumber\\
&& B_{y}=\tfrac{B_0}{L_{0}}((2a-c)yz-(1+La)y+bxz+\tfrac{1}{2} j_{sep}x), \nonumber\\
&& B_{z}=\tfrac{B_0}{L_{0}}(a(Lz-z^2)+\tfrac{1}{2} cx^2+(a-\tfrac{1}{2} c)y^2+bxy).\label{eq:magfield1}
\end{eqnarray}
The details of how this analytical field was formed can be found in Appendix~\ref{sec:app}. With a suitable choice of the parameters, $a$, $b$, $c$, $L$ and $j_{sep}$ (see Appendix~\ref{sec:app}), this magnetic field contains two 3D nulls: a positive null positioned at (0,0,0) orientated with its fan in a $y$-$z$ plane and a negative null positioned at (0,0,$L$) with its fan in an $x$-$z$ plane. A separator lies between the nulls along the $z$-axis and the electric current associated with this magnetic field is ${\bf {j}} = \tfrac{B_{0}}{\mu L_{0}}(0,0,j_{sep})$. Hence, the current is uniform and runs parallel to the $z$-axis throughout the domain. In all experiments discussed in this paper the scaling factors $B_{0}$ and $L_{0}$ are set to one, $a = 0.5$, $b = 0.75$, $c = 0.25$, $L=1$ and $j_{sep}^2<6$ (see Appendix~\ref{sec:app}). A different value of $j_{sep}$ is imposed in each experiment.
\subsection{Plasma}

All the experiments discussed here have an initial uniform density of $\rho_{0} = 1.5$, an initial internal energy per unit mass of $\epsilon_{0} = 1.5$ and an initial velocity of ${\bf{v_{0}}} = {\bf {0}}$ (where the subscript ``0'' indicates initial normalised values). From the equation of state we know that $p = \rho \epsilon (\gamma - 1)$, which implies that the initial normalised pressure, $p_{0} = 1.5$. 

\begin{figure}
\centering
\includegraphics[width=0.36\textwidth,clip]{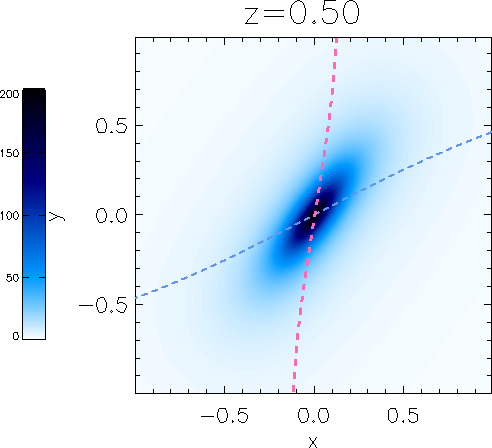}
\caption{Contour plot of the initial plasma beta in a cut perpendicular to the $z$-axis (separator) at $z=0.5$ for the main experiment with $j_{sep}=1.5$. Over plotted are the intersections of the lower and upper null's separatrix surfaces with the $z=0.5$ plane in the initial state (pale-blue/pink dashed lines, respectively).}\label{fig:j1.5_beta}
\end{figure}

Although the pressure is uniform throughout the domain, the magnetic field strength varies. Initially the mean plasma beta in the domain is $\beta = 7.8$. Half way along the separator (at $x = y = 0$, $z = 0.5$), due to the close proximity to the nulls where the plasma beta is infinite, the plasma beta is high, $\beta = 192$. Fig.~\ref{fig:j1.5_beta} displays contours of the plasma beta in a cut across the separator and fans at $z=0.5$ in the initial state. The plasma beta is large in the vicinity of the separator (at $x=y=0$) but falls off rapidly away from this region. Separators embedded in regions with such high plasma betas are either cluster separators which connect two null points within a null cluster \citep{Parnell10b} or are separators in planetary magnetospheres \citep[e.g.][]{Dorelli07}. Plasma betas of between 1 and 10 have been determined in the Earth's magnetosphere, but obviously these will be much higher near null points, \citep{Trenchi2008}.   

The value of our plasma beta is high due to the value of the strength of the initial pressure. This high pressure ensures only two nulls exist in the model during the entire relaxation process. It is possible to achieve a lower plasma beta by either increasing the initial length of the separator, $L$, or increasing the magnetic field scaling factor, $B_{0}$, but both methods can lead to an increase in numbers of nulls soon after the relaxation begins. Alternative scenarios will be considered to seek a low-plasma beta separator current layer in future work.

In this paper, we have normalised the times to the fast-mode crossing time along the separator, from one null to the other, as follows. The fast-mode crossing time is given by
$$ t_f = \int_{0}^{L} \frac{1}{\sqrt{c_s^2+c_A^2(z)}}dz, $$
where $c_{s}$ is the sound speed ($\sqrt{\epsilon\gamma(\gamma - 1)}$) and $c_{A}(z)$ is the Alfv\'en speed ($\sqrt{B(z)^{2}/\rho} = B_0(a(Lz -z^2))/(L_0\sqrt{\rho}) $).
Initially, the sound speed is uniform throughout the domain with $c_s = 5/3$ and the magnetic field along the separator is known analytically, hence, we find the value of $t_f$ integrated along the separator to be $t_f=0.77$ (using the magnetic field parameters detailed previously).

The value of the fast-mode crossing time was also calculated from both nulls along the shortest paths to the domain boundaries. We found the crossing times from the lower null to the nearest $x$, $y$ and $z$ boundaries are $t_f=0.71$, $t_f=0.67$ and $t_f=0.74$, respectively. Similarly, for the upper null, the fast-mode crossing times from this null to the nearest $x$, $y$ and $z$ boundaries are $t_f=0.65$, $t_f=0.74$ and $t_f=0.74$, respectively. 

\subsection{Initial null point properties}
We now look more closely at the initial magnetic field with the parameters set to $B_0=1$, $L_{0}=1$, $a = 0.5$, $b = 0.75$, $c = 0.25$ and $L=1$ as previously stated.  Information about the nature of the field can be determined by considering the eigenvalues and eigenvectors associated with the local (linear) field about each null. The eigenvalues are,
\begin{eqnarray}
&& \lambda_{sl} = -0.25 - \frac{\alpha}{2}, \hspace{0.3cm} \lambda_{su} = 0.25 + \frac{\alpha}{2},\nonumber\\
&& \lambda_{f_{1}l} = 0.5, \hspace{1.25cm} \lambda_{f_{1}u} = -0.5,\\
&& \lambda_{f_{2}l} = -0.25 + \frac{\alpha}{2}, \hspace{0.2cm} \lambda_{f_{2}u} = 0.25 - \frac{\alpha}{2},\nonumber
\label{eq:evals}
\end{eqnarray}
where $\alpha = \sqrt{6.25 - j_{sep}^{2}}$ and the subscripts ``$s,f_{1},f_{2}$'' refer to the spine, minor and major separatrix-surface eigenvalues, respectively, and ``$l,u$'' refer to the lower (positive) and upper (negative) nulls, respectively.  The eigenvectors associated with these eigenvalues are
\begin{eqnarray}
&& {\bf{e}}_{sl} = \Bigg(\frac{j_{sep}}{2.5+\alpha},1,0\Bigg)^{T}, \hspace{0.2cm} {\bf{e}}_{su} = \Bigg(\frac{4+2\alpha}{3+2j_{sep}},1,0\Bigg)^{T},\nonumber\\
&& {\bf{e}}_{f_{1}l} = \Bigg(0,0,1\Bigg)^{T}, \hspace{1.1cm} {\bf{e}}_{f_{1}u} = \Bigg(0,0,1\Bigg)^{T},\\
&& {\bf{e}}_{f_{2}l} = \Bigg(\frac{j_{sep}}{2.5-\alpha},1,0\Bigg)^{T},\hspace{0.2cm} {\bf{e}}_{f_{2}u} = \Bigg(\frac{4-2\alpha}{3+2j_{sep}},1,0\Bigg)^{T}.\nonumber 
\label{eq:evecs}
\end{eqnarray}
In this paper, experiments with $j_{sep} = 0.75, 1.0, 1.25, 1.5$ and 1.75 are investigated. All values of $j_{sep}$ are chosen such that both nulls are initially improper radial nulls, i.e., the eigenvalues of the fans are real and distinct ($\lambda_{f_{1}l}$ $\neq$ $\lambda_{f_{2}l}$ and $\lambda_{f_{1}u}$ $\neq$ $\lambda_{f_{2}u}$ - see \citet{Parnell96} for more details on the nature of 3D nulls). 

\begin{figure}[!h]     \begin{center}
   \includegraphics[width=0.475\linewidth]{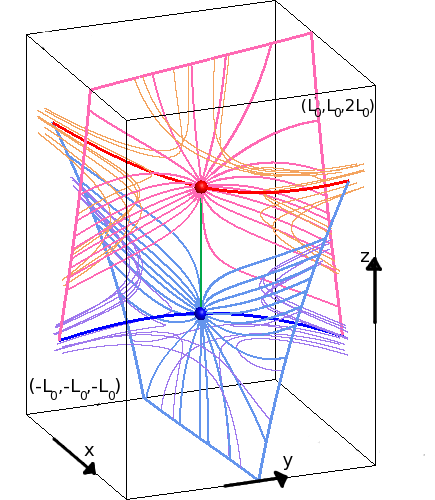}
   \includegraphics[width=0.45\linewidth]{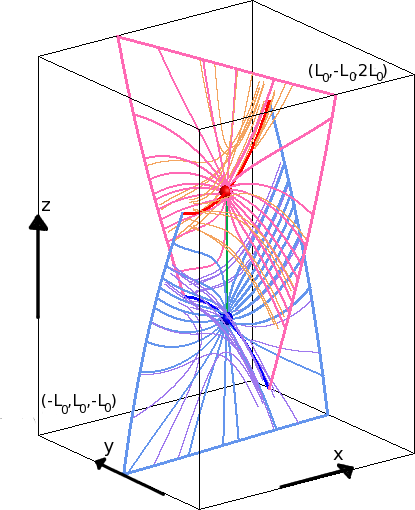}
   \includegraphics[width=0.45\linewidth]{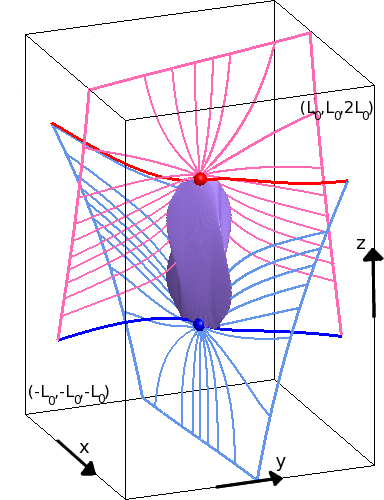}
   \includegraphics[width=0.45\linewidth]{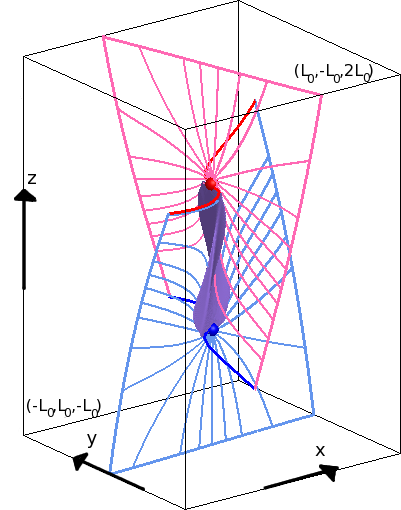}
  \end{center}
    \vspace{-0.325\textwidth}
\centerline{\Large \bf \hspace{0.01 \textwidth}  \color{black}{\small(a)}\hfill}
    \vspace{-0.02\textwidth}
\centerline{\Large \bf \hspace{0.275\textwidth}  \color{black}{\small(b)}\hfill}
    \vspace{0.25\textwidth}
\centerline{\Large \bf \hspace{0.01 \textwidth}  \color{black}{\small(c)}\hfill}
    \vspace{-0.01\textwidth}
\centerline{\Large \bf \hspace{0.275\textwidth}  \color{black}{\small(d)}\hfill}
    \vspace{0.01\textwidth}
\caption{The magnetic skeleton of the initial field (a \& b) and the final equilibrium field (c \& d) viewed from two different angles showing the lower/upper nulls -- blue/red spheres, spines -- thick blue/red lines, separatrix-surface field lines -- pale-blue/pink lines and separator -- thick green line. General field lines are drawn around the lower/upper initial null's spines (lilac/orange). The pale-blue/pink thick lines indicate where the separatrix surfaces from the lower/upper nulls intersect the boundaries of the domain. The equilibrium plots (c \& d) include an isosurface (purple) of the parallel current drawn at $20\%$ of the maximum value in the equilibrium field.}
\label{fig:j1.5_skel}
    \end{figure}

The main case detailed in this paper has $j_{sep} = 1.5$ and contains nulls with the following eigenvalues and eigenvectors:
\begin{eqnarray}
&& \lambda_{sl} = -1.25, \hspace{0.1cm} \lambda_{su} = 1.25,\nonumber\\
&& \lambda_{f_{1}l} = 0.5, \hspace{0.4cm} \lambda_{f_{1}u} = -0.5,\nonumber\\
&& \lambda_{f_{2}l} = 0.75, \hspace{0.2cm} \lambda_{f_{2}u} = -0.75,\nonumber\\
&& {\bf{e}}_{sl} = \Bigg(\frac{1}{3},1,0\Bigg)^{T}, \hspace{0.1cm} {\bf{e}}_{su} = \Bigg(\frac{4}{3},1,0\Bigg)^{T},\\
&& {\bf{e}}_{f_{1}l} = \Bigg(0,0,1\Bigg)^{T}, \hspace{0.1cm} {\bf{e}}_{f_{1}u} = \Bigg(0,0,1\Bigg)^{T},\nonumber\\
&& {\bf{e}}_{f_{2}l} = \Bigg(3,1,0\Bigg)^{T},\hspace{0.1cm} {\bf{e}}_{f_{2}u} = \Bigg(0,1,0\Bigg)^{T}. \nonumber\label{eq:j1.5evalsevecs}
\end{eqnarray}
The magnetic skeleton of this initial configuration (the main experiment described in this paper), which was found using the methods described in \citet{HaynesP07} and \citet{Haynes10}, is shown in Figs.~\ref{fig:j1.5_skel}a and \ref{fig:j1.5_skel}b. The separator (green) linking the two nulls is formed from the intersection of the two separatrix surfaces (pink or pale-blue field lines). These surfaces are seen to twist gently about the $x=y=0$ line (i.e. the separator), thus the angle between the two separatrix surfaces varies along the separator.
\section{Results}\label{sec:results}
\begin{figure}
\centering
\includegraphics[width=0.4\textwidth,clip]{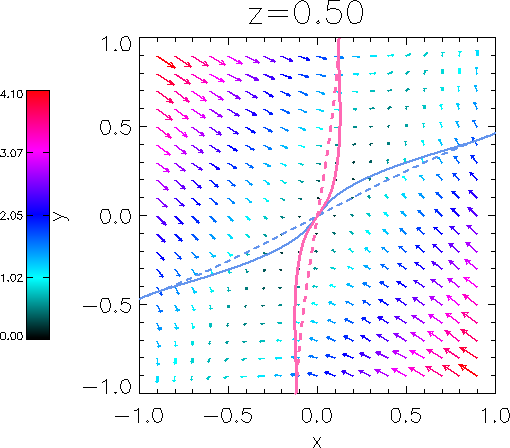}
\caption{Arrows display the initial Lorentz force in the plane $z = 0.5$ for the main experiment with initial current $j_{sep} = 1.5$. The pale-blue/pink lines indicate where the separatrix surfaces from the lower/upper nulls of the initial (dashed) and equilibrium (solid) magnetic fields intersect the plane. The arrows indicate the strength (by their size and colour) and direction of the Lorentz force in the plane. The length of the arrows has been normalised to the maximum value of $|{\bf {j}} \times {\bf {B}}|$ in the domain.}\label{fig:j1.5_lorarrows}
\end{figure}

There is an initial non-zero Lorentz force in all the initial magnetic fields examined in this paper, which acts in planes perpendicular to the separator, causing the separatrix surfaces to fold towards each other as soon as the relaxation begins. In Fig.~\ref{fig:j1.5_lorarrows}, the size and strength of this force in a plane perpendicular to the separator, mid way along its length, is plotted for the main experiment, where $j_{sep} = 1.5$. 

In this paper, we show that the collapse of the separatrix surfaces about the separator is analogous to the collapse of separatrices about a 2D null point \citep{FF11}, and that a current layer is formed at the separator. However, the collapse also causes gradients to develop in the plasma pressure which provide a counter force slowing the collapse and eventually creating an equilibrium.

Initially there are no pressure gradients to balance the non-zero Lorentz force so, as soon as the experiment starts, waves are launched and the system evolves under non-resistive MHD (i.e., there is no reconnection and so there is no transfer of flux between the four flux domains about the separator). The system relaxes ideally, save for the damping of waves via viscous effects. Since the magnetic field is complex, it is not possible to form an equilibrium without current layers. Current layers form at topological or geometrical features, so in this model they can form at either the 3D null points, the separatrix surfaces, spines or the separator. We find that the MHD relaxation causes current accumulations to form along the separator, on the separatrix surfaces close to the separator (Figs.~\ref{fig:j1.5_skel}c and \ref{fig:j1.5_skel}d) and on the separatrix surfaces close to the boundaries at the top and bottom of the box. The latter form due to the boundary conditions which prevent the separatrix surfaces moving on them. 

The system evolves to what appears to be an equilibrium state, by $t = 51.67t_{f}$ for the main experiment. 
In the rest of this section, we focus on this experiment where $j_{sep} = 1.5$. We consider the structure of the magnetic skeleton of the final equilibrium field and consider the appearance of the current accumulation (Sect.~\ref{sec:skel}), the evolution of the energetics during the relaxation (Sect.~\ref{sec:energy}) and the details of the force balance of the final equilibrium (Sect.~\ref{sec:tot}).
\subsection{Magnetic skeleton}\label{sec:skel}
\begin{figure}
\centering
\includegraphics[width=0.4\textwidth,clip]{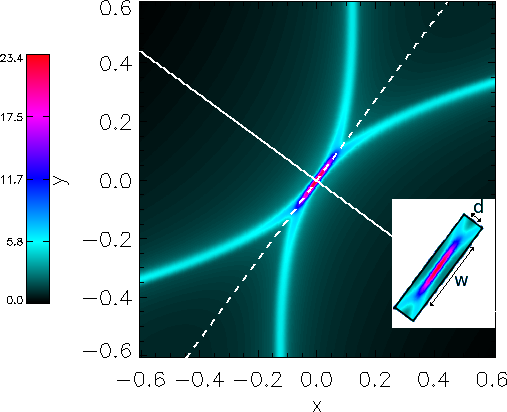}
\caption{A cut through the skeleton of the final magnetic field in the plane $z=0.5$ perpendicular to the separator with contours of $|{\textbf{j}}|$ showing the current layer at the centre. The insert shows the separator current layer's depth (labelled $d$) and width (labelled $w$). On the main plot, lines which cut across the width (dashed, white) and through the depth (solid, white) of the current layer are shown.  The length, $l_{sep}$, of the current layer (not shown) is directed out of the plane of this cut and is equal to the length of the separator in the final equilibrium.} \label{fig:j1.5_2Dskel}
\end{figure}
The structure of the final equilibrium magnetic field is described by its magnetic skeleton \citep{HaynesP07,Haynes10}. In Figs.~\ref{fig:j1.5_skel}c and \ref{fig:j1.5_skel}d the equilibrium field's magnetic skeleton is shown along with the current layer that has formed (purple isosurface of the current parallel to the magnetic field, $j_{\parallel}$).  It is clear that a current layer has been created along the separator (the details of which are discussed in Sect.~\ref{sec:current}). {The profile of this current layer in the $z=0.5$ plane perpendicular to the separator is shown in Fig.~\ref{fig:j1.5_2Dskel}. Contours of $|{\textbf{j}}|$ highlight the strong current layer about the separator and the enhanced current along the separatrix surfaces. Everywhere else the current is zero indicating that the current layer is embedded in a potential magnetic field. The solid and dashed white lines here are plotted through the depth and across the width of the current layer, respectively, in this cut at $z=0.5$.}

During the whole relaxation process only two nulls are found in each time step and the topology of the system does not change (which implies that no numerical dissipation has occurred). At the start of the relaxation the nulls move slightly further away from each other along the $z$-axis, but then come back towards each other briefly before slowly moving apart along the $z$-axis towards the end of the relaxation. The rate of movement is 1.1$\times 10^{-3}\;L_{0}/t_{f}$ just after the initial oscillations die down and 4.3$\times 10^{-4}\; L_{0}/t_{f}$ at the end of the relaxation. This very slow continuous lengthening of the separator, after the system appears close to equilibrium, suggests that the system is asymptotically approaching an equilibrium, as is seen in the formation of current layers at 2D nulls \citep[e.g.,][]{Klapper98,Craig05,FF11} and 3D nulls \citep[e.g.,][]{FF12,FF13}. This asymptotic behaviour is considered in more detail in Sect.~\ref{sec:growth}. 

From the 3D images seen in Fig.~\ref{fig:j1.5_skel} the spines and separatrix surfaces associated with these nulls do not appear to have changed greatly between the initial state and final equilibrium, because they are line tied at the boundaries. However, we know that the current has changed considerably within the domain since initially the current is uniformly distributed, but in the equilibrium state it has accumulated about the separator and, therefore, the magnetic field must have changed. In order to visualise the changes in the magnetic field, we have plotted (along with the Lorentz force which has already been discussed) the magnetic skeletons of the initial and equilibrium fields in a 2D cut at $z=0.5$ in Fig.~\ref{fig:j1.5_lorarrows}. This cut reveals that the separatrix surfaces have warped from their original fairly straight formation (dashed pale-blue/pink lines) to lines that now have a point of inflection at the separator such that they run almost concurrently in the vicinity of the separator (solid pale-blue/pink lines). Indeed, this 2D cut perpendicular to the separator reveals that the separatrix surfaces form cusps exactly like those seen in the collapse of the magnetic field about a 2D null \citep[e.g.,][]{Craig05, Pontin05,FF11}. The cusp regions form due to the nature of the pressure which, initially uniform, is changed through the relaxation. This is discussed in Sect.~\ref{subsec:pressure}.

From the isosurfaces of the current layer shown in Figs.~\ref{fig:j1.5_skel}c and \ref{fig:j1.5_skel}d we can see a number of interesting characteristics including the fact that it is twisted and that it has the beginnings of "wing-like" features where the current enhancement extends out along one or both separatrix surfaces. These extended enhancements seen along the separatrix surfaces were also found in the current layers formed from the collapse of a 2D null \citep[e.g.,][]{FF11}. The isosurface of current seen here is similar in shape to that of a hyperbolic flux tube about a quasi-separator as described by \citet{TGN03}: its cross-sections in cuts perpendicular to the separator start essentially as thin ellipses at one end aligned with the separatrix surface of the nearest null, then become X-like with narrow enhanced current wings along both separatrix surfaces, before returning to thin ellipses aligned with the separatrix surface of the null at the other end.  In Sect.~\ref{sec:current}, the characteristics of the current layer are studied in detail. 

\begin{figure}
\centering
\includegraphics[width=0.92\linewidth,clip]{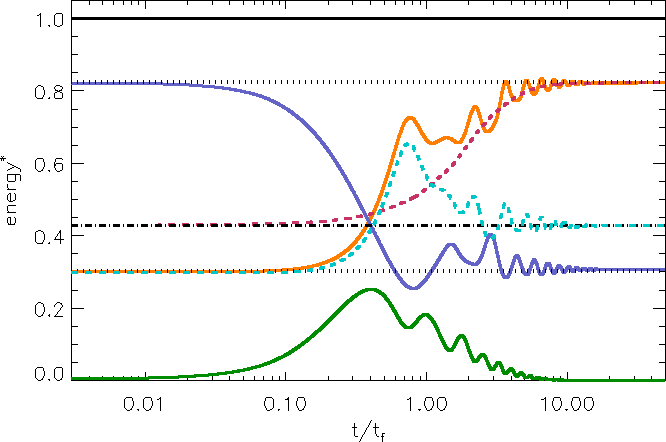}
\caption{Kinetic (green), magnetic (blue), internal (orange) and total (black solid) energies along with the adiabatic (cyan dashed) and viscous heating terms (red dashed) for the experiment with initial current $j_{sep} = 1.5$. Values of 10.849, 26.786 and 37.764 have been subtracted from the magnetic, internal and total energies and -0.339 and 26.697 have been subtracted from the viscous heating and adiabatic terms, respectively, for representational purposes.}\label{fig:j1.5_energy}
\end{figure}
\subsection{Energetics}\label{sec:energy}

Fig.~\ref{fig:j1.5_energy} displays the kinetic, magnetic, internal and total energies {along with the cumulative viscous heating and adiabatic terms (dashed lines) integrated over the whole 3D domain as a function of time. All energies, except the kinetic energy, have been shifted on the $y$-axis for representational purposes. In particular, the internal and magnetic energies have been moved such that the initial internal energy is plotted at the same point on the $y$-axis as the final magnetic energy (lower dotted line). The cumulative adiabatic heating curve also starts from this same point, whilst the cumulative viscous heating curve starts from the point on the $y$-axis where the shifted cumulative adiabatic heating curve ends (dot-dashed line). The upper dotted line indicates the value of the shifted initial magnetic energy in this plot which coincides with the value of the final internal energy.

As required by the closed boundary conditions, the total energy is conserved throughout the run, with a standard deviation of just 0.002\% of the mean, indicating that any energy losses through the boundaries or via numerical dissipation during the relaxation are negligible.

Looking more closely at the energy curves we can see that the difference between the initial magnetic energy and the final magnetic energy is the same as the difference between the initial and final internal energies (as indicated by the dotted lines on Fig.~\ref{fig:j1.5_energy}), which they must be since the initial and final kinetic energies are zero. The conversion of energy from magnetic to internal (via kinetic energy) occurs through one of two processes adiabatic or viscous heating (Eq.~\ref{eq:inte}). In Fig.~\ref{fig:j1.5_energy}, we show that the sum of the cumulative adiabatic and cumulative viscous heating terms equal (to within numerical error) this change in magnetic/internal energy. The fact that these two heating terms can account for all the magnetic energy lost during the relaxation indicates that any magnetic reconnection, caused by numerical diffusion in the system, is negligible. 

As we have already said, the collapse of the initial state, which is not in force balance, creates fast magnetoacoustic waves and, hence, kinetic energy. As these waves bounce across the box they compress or expand the plasma giving rise to either adiabatic heating or cooling, respectively. The oscillatory behaviour in the magnetic, kinetic and internal energies, as well as the cumulative adiabatic heating term are clear signatures of this behaviour. (Note, the periods of these oscillations confirm that the waves in the system are fast magnetoacoustic waves). At the same time, due to the presence of viscosity within the system, these waves are damped giving rise to viscous heating. The cumulative viscous heating term monotonically increases since viscosity only acts to reduce the amplitude of waves and, hence, only converts kinetic energy into internal energy and not vice-versa.
   
The increase in internal energy comes mostly from viscous heating, which is three times bigger than the adiabatic heating. This indicates that the relaxation process is dominated by viscous damping. By considering experiments with identical initial setups, but with different viscosities, it is possible to show that the initial and final internal and magnetic energies in these systems are the same (to within numerical error), however, the proportion of viscous heating to adiabatic heating is greater in the experiment with high viscosity indicating that increasing the viscosity increases the rate of damping, but does not effect the final equilibrium state. This is in agreement with \citet{FF11} and \citet{FFthesis} who both analytically (in 1D \& 2D) and numerically (in 1D, 2D \& 3D) demonstrated that the final equilibria of non-resistive MHD relaxation processes principally depend on the differences between the final and initial total pressures in the system.

By $t = 20t_{f}$, the oscillations in all the energies are basically completely damped. After this the energies maintain constant values, indicating the system has essentially achieved an equilibrium state. 
\begin{figure}[!h]
 \centerline{\includegraphics[width=0.75\linewidth]{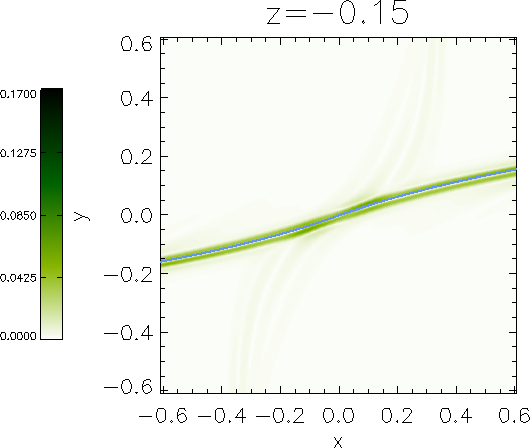}}
 \centerline{\includegraphics[width=0.75\linewidth]{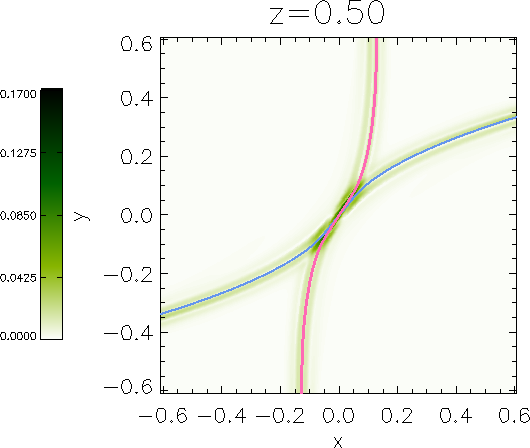}}
 \centerline{\includegraphics[width=0.75\linewidth]{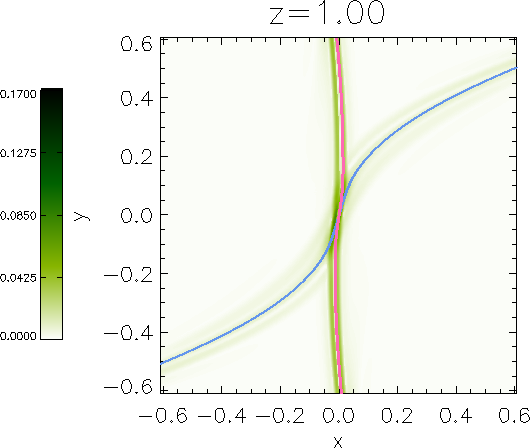}}
    \vspace{-0.675\textwidth}
\centerline{\Large \bf \hspace{0.03 \textwidth}  \color{black}{\small(a)}\hfill}
    \vspace{0.3\textwidth}
\centerline{\Large \bf \hspace{0.03 \textwidth}  \color{black}{\small(b)}\hfill}
    \vspace{0.28\textwidth}
\centerline{\Large \bf \hspace{0.03 \textwidth}  \color{black}{\small(c)}\hfill}
    \vspace{0.01\textwidth}
\caption{Contours of the total force in the final equilibrium state in planes $z =$ (a) -0.15 (below the lower null), (b) 0.5 (through the separator) and (c) 1.0 (through the separator, very close to the upper null). The pale-blue/pink lines indicate where the separatrix surfaces from the lower/upper nulls intersect the plane.}
\label{fig:j1.5_totalforce}
    \end{figure}
\begin{figure}[!h]
\centerline{\includegraphics[width=1.0\linewidth]{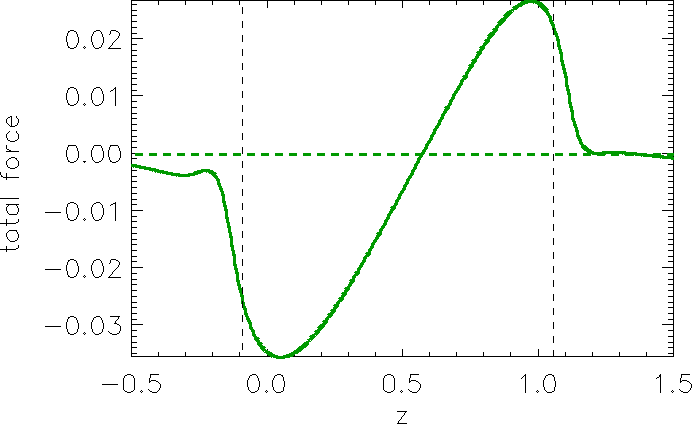}}
\centerline{\includegraphics[width=1.0\linewidth]{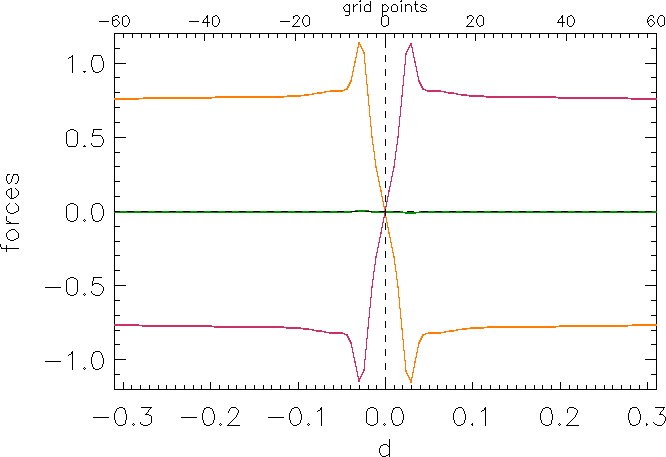}}
\centerline{\includegraphics[width=1.0\linewidth]{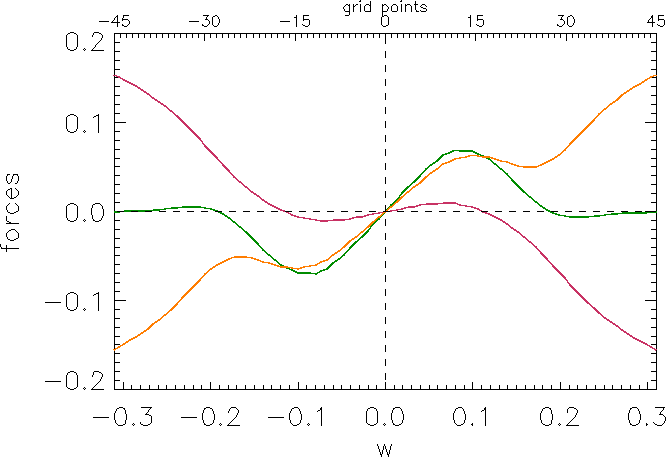}}
    \vspace{-0.7\textwidth}
\centerline{\Large \bf \hspace{0.03 \textwidth}  \color{black}{\small(a)}\hfill}
    \vspace{0.325\textwidth}
\centerline{\Large \bf \hspace{0.03 \textwidth}  \color{black}{\small(b)}\hfill}
    \vspace{0.3\textwidth}
\centerline{\Large \bf \hspace{0.03 \textwidth}  \color{black}{\small(c)}\hfill}
    \vspace{0.01\textwidth}
\caption{(a) The $z$ component of the total force (solid) along the $z$-axis which, here, equals the plasma-pressure force and the perpendicular component of the total force (dashed) along the $z$ axis. Black dashed lines highlight the positions of the nulls. Plots (b) through the depth  and (c) across the width of the current layer in the plane $z = 0.5$ showing the Lorentz force -- orange, plasma-pressure force -- red and the total force -- green. Black dashed lines highlight the position of the separator in (b) and (c).}
\label{fig:j1.5_forces}
   \end{figure}

Finally, we note here that the total integrated current in the domain is initially $18.0$, but it falls to $13.5$ during the relaxation. This fall in current is simply a consequence of the fact that, on the boundaries, the magnetic field parallel to the boundaries may vary and, hence, this is not unexpected.

\subsection{Total force}\label{sec:tot}
To check in more detail that our final state is an equilibrium, we first consider the balance of the Lorentz force and the plasma-pressure force. Filled contours of the total force (the Lorentz force, ${\bf j}\times{\bf B}$, plus the plasma-pressure force, $-\nabla p$), drawn in three different planes perpendicular to the separator, reveal that the total force in the final state is zero everywhere, except very close to the separator and along the separatrix surface of the nearest null to the plane plotted (Fig.~\ref{fig:j1.5_totalforce}). The lack of force-balance in the immediate vicinity of these topological features is not surprising since similar behaviour is found in the equilibrium field associated with collapsed 2D and 3D null points where an infinite-time collapse of the null points is seen \citep[e.g.,][]{Klapper98,Craig05,FF11,FF12,FF13}. Thus, these highly-localised, residual forces suggest that separators may also undergo an infinite-time collapse. Further evidence of this is given in Sect.~\ref{sec:growth}. 

Along the separator itself the Lorentz force vanishes (since ${\bf j}$ remains parallel to the $z$-axis along the separator) and so the total force here is simply the pressure force, Fig.~\ref{fig:j1.5_forces}a. It acts outwards from around the middle of the separator towards the nulls and is small outwith the separator along the $z$-axis. 

In a 1D cut through the depth (e.g. the solid white line in Fig.~\ref{fig:j1.5_2Dskel}) and across the width (e.g. the dashed white line in Fig.~\ref{fig:j1.5_2Dskel}) of the current layer, in the plane $z = 0.5$, the Lorentz and pressure forces behave similarly, but are opposite in sign. This means the total force vanishes everywhere except where it crosses the current layer, Figs.~\ref{fig:j1.5_forces}b and \ref{fig:j1.5_forces}c. Note, the residual force through the depth is too weak to be seen in this graph.  These small residual net forces at the current layer indicate that the current here is still growing, as expected in the case of an infinite-time singularity. \cite{FF11} show similar cuts indicating the same sort of behaviour for the total forces through a current layer formed after the collapse of a 2D null. Residual forces for the collapse of a 2D null or a 3D separator are therefore found to lie within or on the edge of the current layer. The net force through the depth of the current layer, which has a peak magnitude of 0.026 (so not visible in Fig.~\ref{fig:j1.5_forces}b), acts to squeeze the current layer thinner. Reconnection will eventually occur at the current layer once it is sufficiently thin such that numerical diffusion becomes important. We stop all experiments discussed in this paper before this takes place. The net force across the width acts to widen the current layer. It has a peak magnitude of 0.071, some 2.5 times larger than the net forces along the length of the current layer and through its depth. This suggests that the current layer is more likely to widen rather than lengthen as the slow relaxation continues.

The second test we carried out to see if our final state is an equilibrium was to check the value of the pressure along the magnetic field lines in the final state. In our system, an equilibrium is achieved when the forces (Lorentz and pressure) balance. Taking the dot product of the sum of these forces with ${\bf B}$ gives
$$\left({\bf j}\times{\bf B}-\nabla p\right)\cdot {\bf B} = -{\bf B}\cdot\nabla p = 0\;.$$
This implies that, in an equilibrium state, pressure will be constant along field lines. Although not plotted here, the pressure was found to remain constant (to within $1.5\%$) along magnetic field lines indicating that, in general, our system may have achieved an equilibrium state.

\section{Nature of the current layer}\label{sec:vary}
So far we have focussed on just one experiment with an initial current of $j_{sep}=1.5$. Here, however, we now consider the effects of varying the magnitude of the initial uniform current $j_{sep}$ on the nature of the current layer formed in the final equilibrium states. In these experiments the initial setup is identical apart from the initial current, $j_{sep}$ which takes one of the following values, 0.75, 1.0, 1.25, 1.5 and 1.75. 

First, the magnitude of the current within and outside the current layer is discussed in Sect.~\ref{sec:current}. Then the twist of the current layer is described in Sect.~\ref{sec:twist}, whilst in Sect.~\ref{sec:dim} the dimensions of the current layer are calculated. The behaviour of the plasma pressure and the balance of the forces through the current layer are studied in Sect.~\ref{subsec:pressure} and Sect.~\ref{sec:forces}, respectively. Finally, Sect.~\ref{sec:growth} verifies the infinite time collapse of the field about the separator and calculates the growth rate of the current layer.

\subsection{Current intensity}\label{sec:current}
\begin{figure}[!h]
\centerline{\includegraphics[width=1.0\linewidth]{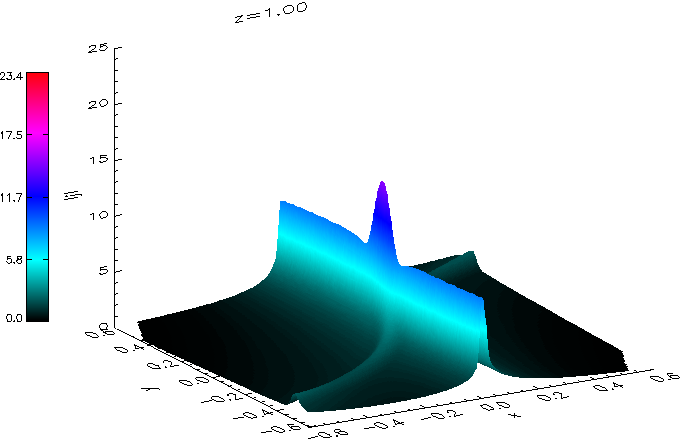}}
\centerline{\includegraphics[width=1.0\linewidth]{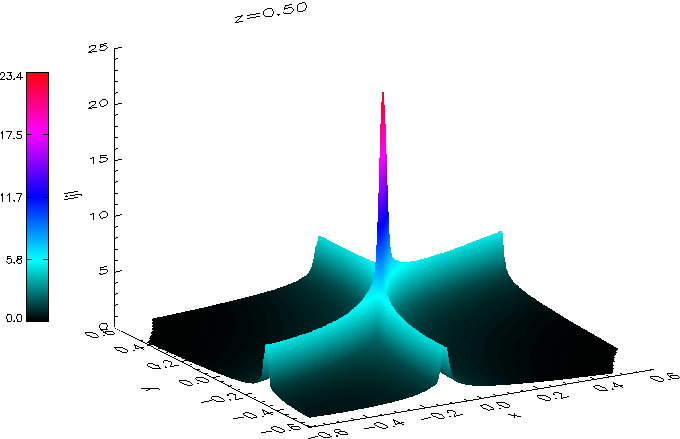}}
\centerline{\includegraphics[width=1.0\linewidth]{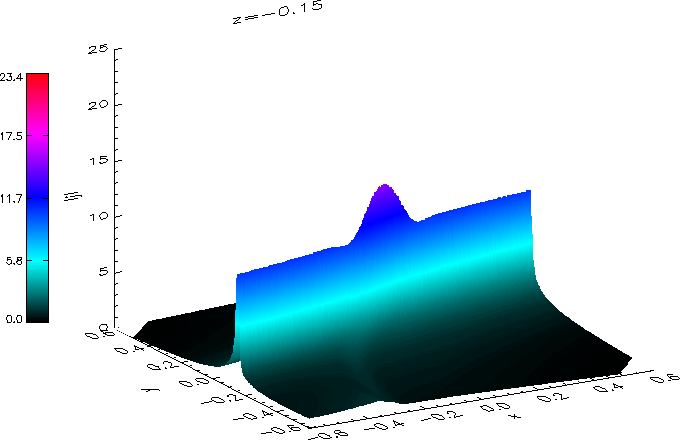}}
  \vspace{-0.65\textwidth}
\centerline{\Large \bf \hspace{0.03 \textwidth}  \color{black}{\small(a)}\hfill}
    \vspace{0.31\textwidth}
\centerline{\Large \bf \hspace{0.03 \textwidth}  \color{black}{\small(b)}\hfill}
    \vspace{0.28\textwidth}
\centerline{\Large \bf \hspace{0.03 \textwidth}  \color{black}{\small(c)}\hfill}
    \vspace{0.01\textwidth}
\caption{Surface plots of $|{\bf {j}}|$ in the final equilibrium in the planes $z =$ (a) 1.0, (b) 0.5 and (c) -0.15 for the experiment with initial current $j_{sep} = 1.5$.}
\label{fig:j1.5_current}
    \end{figure}
\begin{figure}[ht]
\centering
 \includegraphics[width=1.0\linewidth]{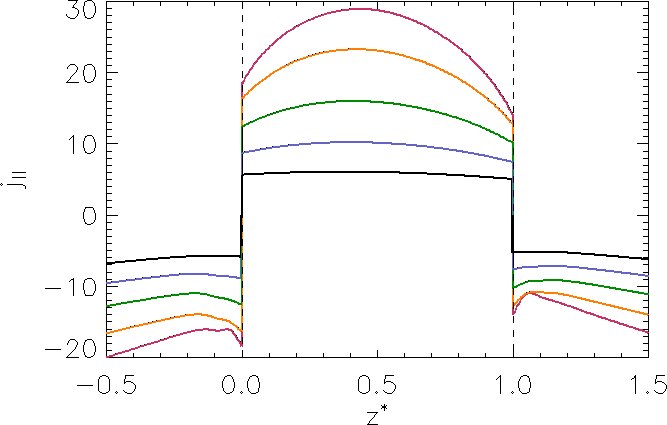}
\caption{$j_{\parallel}$ along the $z$-axis normalised to the length of the respective separator for experiments with initial current $j_{sep} =$ 0.75 (black), 1.00 (blue), 1.25 (green), 1.5 (orange) and 1.75 (red).}
\label{fig:current}
    \end{figure}
    In the final equilibria of all experiments $|{\bf{j}}|$ is found to be strongest along the separator, although enhanced current is also found on the separatrix surfaces (Fig.~\ref{fig:j1.5_current}).
    Everywhere else it is very close to zero. 
    The surface plots of $|{\bf{j}}|$ in Fig.~\ref{fig:j1.5_current} show the distribution of current in various horizontal planes for the final equilibrium of the main ($j_{sep} = 1.5$) experiment. Fig.~\ref{fig:j1.5_current}a shows a plane perpendicular to the separator just below the upper null. The current in this cut peaks at the separator, but is also strong along the separatrix surface of the upper null. A small enhancement of current along the separatrix surface of the lower null also occurs. In the plane $z=0.5$ (Fig.~\ref{fig:j1.5_current}b), a large, sharp peak of current exists at the separator clearly denoting the position of the current layer. The locations of both separatrix surfaces are also clearly visible with ridges of current, approximately 4.2 times smaller than that at the separator current layer, along them. In Fig.~\ref{fig:j1.5_current}c,  the current in a plane just below the lower null, and therefore not cutting the separator, is plotted. There is an enhanced ridge of current all the way along the separatrix surface of the lower null which peaks on the $z$-axis. This cut suggests that the current layer may extend beyond the separator. These plots only show the magnitude of the current. Later we consider the direction of the current within the current layer.

The ratio of the mean current in the separator current layer about the separator over the mean current on the separatrix surfaces ($j_{cl}/j_{ss}$) increases with the initial current $j_{sep}$ from a factor of just 2.6 when $j_{sep} = 0.75$ up to 3.7 for the case with $j_{sep} = 1.75$ (Table~\ref{tab:vals}). These factors are all much greater than two indicating that the current at the separator is not simply a combination of the current enhancements from the two separatrix surfaces, but is itself a genuine current layer associated with the separator.

In Fig.~\ref{fig:current}, the distribution of the parallel current along the $z$-axis is plotted. The final lengths of the separators are all dependent on the initial $j_{sep}$ (see Sect.~\ref{sec:dim} for further details) and so to enable the parallel currents to be compared, the lengths of the separators have all been normalised to one. Thus, in this plot $z^* = (z - z_{ln})/l_{sep}$ where $z_{ln}$ is the $z$ coordinate of the lower null in the final equilibrium and $l_{sep}$ is the length of the separator in the final equilibrium. 
The parallel current ($j_{\parallel}$) along the $z$-axis is positive along the separator, but drops sharply at the nulls becoming negative in sign outside the separator. These negative values increase slightly before decreasing away from the separator. The strong currents at the top and bottom boundaries are a result of the line-tied boundary conditions on the magnetic field which prevent the separatrix surfaces from moving. The local peak in magnitude of these currents just outside the separator, in the experiments with the largest initial currents, suggest that the separator current layers have reverse currents at their ends.  Although not commonly discussed, reverse currents have also been found associated with current layers formed at 2D null points \citep[e.g.,][]{titov93b,Bungey95}. 

The plot of the parallel current along the $z$-axis (Fig.~\ref{fig:current}) has an asymmetric profile, with a greater value as you approach the lower null along the separator than as you approach the upper null, in all experiments. We suspect this is due to asymmetries in the initial field and plan to investigate this further in future work.
Furthermore, the current peaks about 42\%-43\% of the way along the separator in all cases. 
From Fig.~\ref{fig:current}, it is clear that the average and maximum values of $|{\bf {j}}|$ along the separator increase with initial current $j_{sep}$, (as indicated in Table \ref{tab:vals}). The gradient between consecutive maximum and average values increases with $j_{sep}$, except between $j_{sep} =$ 1.5 and 1.75 where the gradient decreases slightly. This is probably due to the fact that the experiment with $j_{sep}=1.75$ was not run for as long as the other experiments, and so is not quite as relaxed. The run was ended early since numerical dissipation, evidenced by the formation of additional nulls, started shortly after the final equilibrium state shown here. It is also noted that in the experiments $|{\bf {j}}|$ along the separator varies more as $j_{sep}$ is increased, since the average of $|{\bf {j}}|$ pulls away from the maximum $|{\bf {j}}|$ (again, as seen in Fig.~\ref{fig:current}). 
\begin{center}
\begin{table*}[ht]
\centering
{\small
\hfill{}
\begin{tabular}{ c c c c c c c c c c }
\hline
$j_{sep}$ & $j_{cl}/j_{ss}$ & Average $|j_{cl}|$ & Maximum $|j_{cl}|$ & Twist & Length $l_{sep}$ & \multicolumn{2}{c}{Depth} & \multicolumn{2}{c}{Width} \\ \hline
          &           &   &   &           &                 &    CM      &   FWHM       &  CM              &   FWHM           \\ \hline
0.75 &  2.6 & 5.90 & 6.18 & 0.111$\pi$ & 1.06 & 0.0226 & 0.0282  & 0.099 & 0.109 \\
1.0  & 3.0 & 9.55 & 10.32 & 0.146$\pi$ & 1.08 & 0.0242 & 0.0219  & 0.155 & 0.136\\
1.25 & 3.4 & 14.48 & 16.10 & 0.203$\pi$ & 1.11 & 0.0243 & 0.0221 & 0.216 & 0.169 \\
1.5 & 3.7 & 20.50 & 23.36  & 0.245$\pi$ & 1.15 & 0.0253 & 0.0199 & 0.309 & 0.223\\
1.75 & 3.7& 25.03 & 29.01  & 0.243$\pi$ & 1.20 & 0.0288 & 0.0210 & 0.428 & 0.305 \\
\end{tabular}}
\hfill{}
\captionsetup{justification=centering}
\caption{Equilibrium parametric values for all five relaxation experiments.}
\label{tab:vals}
\end{table*}
\end{center}

\subsection{Current layer twist}\label{sec:twist} 
From the isosurface of $j_{\parallel}$ in Figs.~\ref{fig:j1.5_skel}c and \ref{fig:j1.5_skel}d and from the contours of current in cuts through the separator in Fig.~\ref{fig:j1.5_current}, we can see that the current layer is twisted, i.e., as $z$ varies, the current layer rotates. Here, we consider how this twist varies with $j_{sep}$, after briefly explaining why such a twist arises. 

Initially, the two separatrix surfaces lie in vertical planes which intersect at an angle dependent on the initial $j_{sep}$ (in the $j_{sep}=1.5$ case the angle is roughly $\pi/3$). Also, the spine's lines from each null, which bound on one edge the separatrix surfaces of the other null, initially lie in $xy$-planes, thus they are at right angles to the initial uniform current. The relaxation process causes the two separatrix surfaces to close up and run almost concurrently in the local vicinity of the separator. Midway along the separator ($z=0.5$) this is achieved by both separatrix surfaces curving equally in towards each other (Fig.~\ref{fig:j1.5_lorarrows}), but at the end of the separator the separatrix surface associated with the local null does not move, instead the other separatrix surface (and thus the spine of the local null) moves. This is due to the initial Lorentz force which, in the $z=0.5$ plane, is such that both separatrix surfaces close in towards each other (see Fig.~\ref{fig:j1.5_lorarrows}). However, at each null the initial Lorentz force is greater across its spine than it is across its separatrix surface. So, at the ends of the separator, the local separatrix surfaces essentially maintain their original positions and thus the current layer must rotate along its length through an angle approximately equivalent to that between the planes of the initial separatrix surfaces. 
Thus the angle, $\theta$, through which the current layer twists between the lower and upper nulls depends on the initial current $j_{sep}$ (Table ~\ref{tab:vals}).
\subsection{Current layer dimensions}\label{sec:dim}
\subsubsection{Length of current layer}

In order to determine the dimensions of the current layer, we need to define where it starts and ends. The length of the current layer, $l_{sep}$, is defined as the distance between the two null points (i.e., the length of the separator) in the final equilibrium. In Sect.~\ref{sec:current}, we have seen that these are  also the points at which the current changes sign. This means we do not include the reverse current regions when determining the length of the current layer. 

During the relaxation the null points move apart along the $z$-axis (as discussed in Sect.~\ref{sec:skel}) and so all the equilibrium current layers have lengths greater than 1 (Table~\ref{tab:vals}). As $j_{sep}$ increases the length of the current layer increases due to the greater initial Lorentz force. 
\begin{figure}[!h]
\centerline{\includegraphics[width=1.0\linewidth]{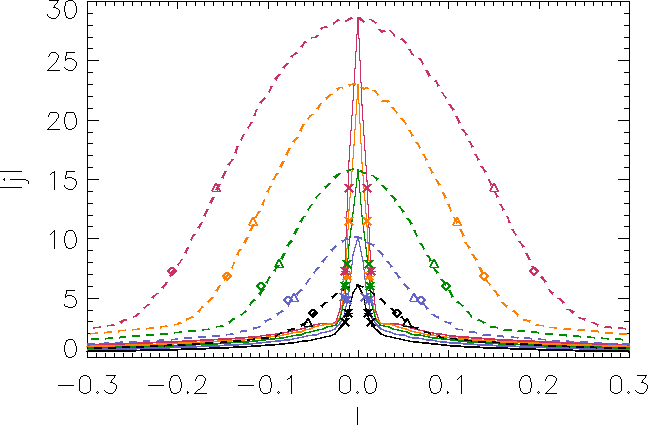}}
\centerline{\includegraphics[width=1.\linewidth]{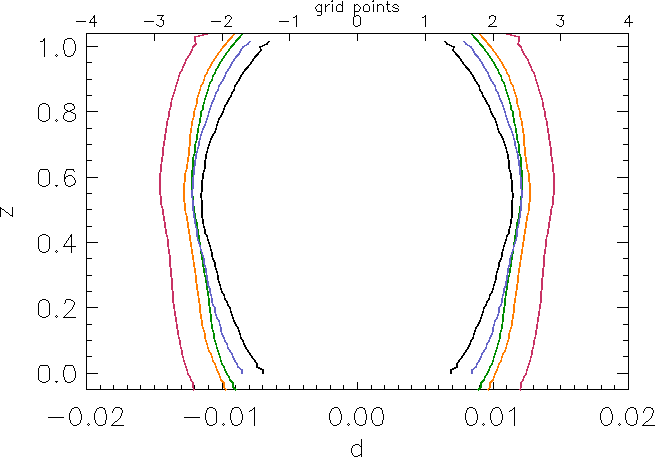}}
\centerline{\includegraphics[width=1.0\linewidth]{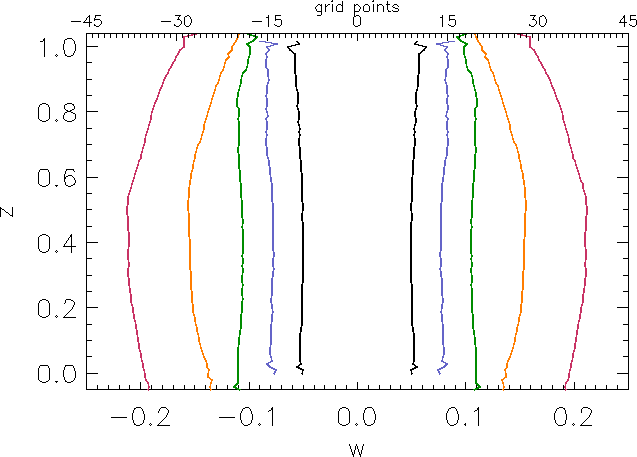}}
    \vspace{-0.725\textwidth}
\centerline{\Large \bf \hspace{0.03 \textwidth}  \color{black}{\small(a)}\hfill}
    \vspace{0.33\textwidth}
\centerline{\Large \bf \hspace{0.03 \textwidth}  \color{black}{\small(b)}\hfill}
    \vspace{0.32\textwidth}
\centerline{\Large \bf \hspace{0.03 \textwidth}  \color{black}{\small(c)}\hfill}
    \vspace{0.01\textwidth}
\caption{(a) $|{\bf {j}}|$ in a cut through the depth (solid) and across the width (dashed) of the current layer in the plane $z = 0.5$ with asterisks and diamonds representing the values of the current contour used to determine the depth and the width of the current layer at this cut for the contour method and the crosses and triangles for the FWHM method, respectively. (b) Outline of the depth and (c) width of the current layer. The colours represent initial current $j_{sep} =$ 0.75 (black), 1.00 (blue), 1.25 (green), 1.5 (orange) and 1.75 (red).}
\label{fig:allj_clplots}
    \end{figure}
\subsubsection{Width and depth of current layer}
By looking at Figs.~\ref{fig:j1.5_skel}  and~\ref{fig:j1.5_2Dskel}, we can see that the current layer's depth is many times smaller than its width which is much shorter than its length. However, quantifying the width and depth of the current layer is not trivial since the current gradually decreases rather than abruptly stops. We consider two approaches to determine the width and depth of the current layer in cuts perpendicular to the separator. The two methods are (i) the contour method which uses the last elliptical current contour, before the current contours deform as they start to extend along the separatrix surfaces (i.e., become X or bone shaped) and (ii) the full width at half maximum (FWHM) of the current. The first method described here is our preferred method because it typically accounts for more of the current about the separator and the values of the current contours vary less between cuts than the second method, but we include both for completeness.

Fig.~\ref{fig:allj_clplots}a shows 1D slices of $|{\bf {j}}|$, in the $z=0.5$ plane, through the depth (solid) and across the width (dashed) of the current layer for all the different experiments. The 1D slices of $|{\bf {j}}|$ through the current layer depth show significantly enhanced $|{\bf {j}}|$ forming a narrow peak about the separator. Elsewhere along this slice the current is small. 

The width and depth of the current layer vary along the current layer's length. Using the contour method, they are defined by examining contours of $|{\textbf{j}}|$ in cuts across the current layer. Plotting a contour in each cut at a value of $|{\textbf{j}}|$ which only outlines the current layer, and not the enhanced current along the separatrix surfaces, allows the width and depth to be measured. In other words, we count only the current down to the inflection point of $|{\textbf{j}}|$ to pick out the current layer (the transition point between elliptical and X-shaped current contours). In all cases, the same contour goes through the two inflection points that lie either side of the separator. Once the correct contour has been found the width and depth of the current layer, along the length of the separator, are determined. 

In Figs.~\ref{fig:allj_clplots}b and \ref{fig:allj_clplots}c, the current layer's depth and width, respectively, determined using the contour method, are plotted, against $z$, for all the different experiments. The current layer depths are greatest away from the nulls and narrowest at either end of the current layers near the nulls (Fig.~\ref{fig:allj_clplots}b). For the current layers with the largest initial $j_{sep}$, the widths follow a similar profile in which they bulge at their middle, but the widths of the other current layers remain essentially constant along the current layer's length (Fig.~\ref{fig:allj_clplots}c).

In order to see how the widths and depths determined using the contour method compare to those calculated with the FWHM method, we determine the depths and widths in the $z = 0.5$ plane using both methods. For each experiment, the results from the two methods (where the contour method is denoted by CM) are presented in Table \ref{tab:vals}. The values of the current contours used to make these measurements in both the contour method and the FWHM method are indicated on the cuts in Fig.~\ref{fig:allj_clplots}a. In general, the FWHM estimates of the current layer's width and depth are smaller than the contour method's, except in the case with the lowest initial current. Both methods indicate that as the initial current increases, so do the dimensions of the current layer. 
In contrast to the contour method, the value of FWHM, hence, the contour used to calculate the widths and depths, varies greatly along the length of the separator because the maximum current along the separator changes quite considerably with length along the separator, as shown in Fig.~\ref{fig:current}. We, therefore, do not feel that the FWHM method is as robust as the contour method. However, the FWHM method does indicate that the higher the initial current, $j_{sep}$, the closer the equilibrium current layer appears to be to a singularity, since the depth of the current layer determined using this method decreases with increasing initial current. 

\subsection{Plasma pressure}\label{subsec:pressure}
\begin{figure}
   \centerline{\includegraphics[width=0.9\linewidth]{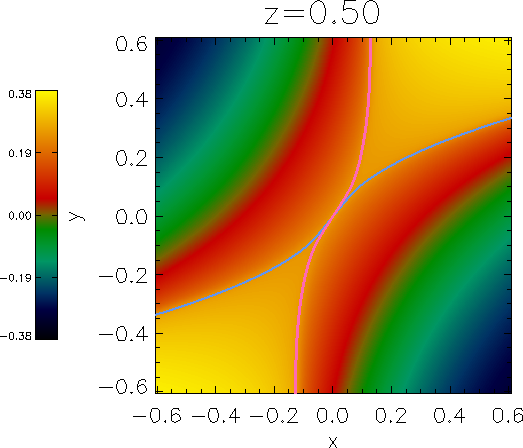}}
   \centerline{\includegraphics[width=0.55\linewidth]{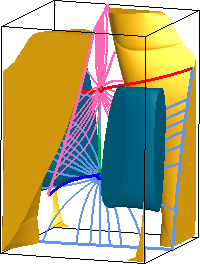}}
   \centerline{\includegraphics[width=0.9\linewidth]{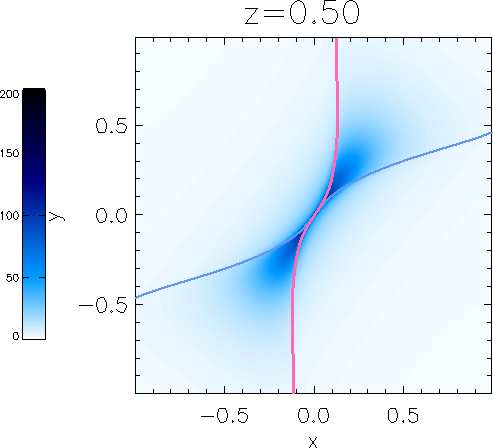}}
    \vspace{-0.8\textwidth}
    \centerline{\Large \bf \hspace{0.03 \textwidth}  \color{black}{\small(a)}\hfill}
    \vspace{0.37\textwidth}
    \centerline{\Large \bf \hspace{0.03 \textwidth}  \color{black}{\small(b)}\hfill}
    \vspace{0.35\textwidth}
    \centerline{\Large \bf \hspace{0.03 \textwidth}  \color{black}{\small(c)}\hfill}
    \vspace{0.01\textwidth}
   \caption{Contours of (a) the pressure difference ($p - p_{0}$) in the $z=0.5$ plane of the final equilibrium and (b) the 3D skeleton of this field with yellow/blue isosurfaces of pressure difference ($p - p_{0}$) drawn at 95\% of the maximum positive/negative value. (c) The plasma beta in the final equilibrium state in the plane $z = 0.5$. The pale-blue/pink lines indicate where the separatrix surfaces from the lower/upper nulls intersect the plane.}
\label{fig:j1.5_pressure}
    \end{figure}
\begin{figure}[!h]
\centerline{\includegraphics[width=0.9\linewidth]{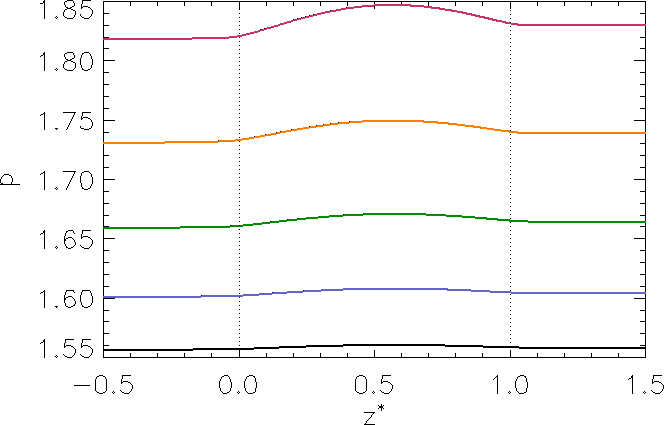}}
\centerline{\includegraphics[width=0.9\linewidth]{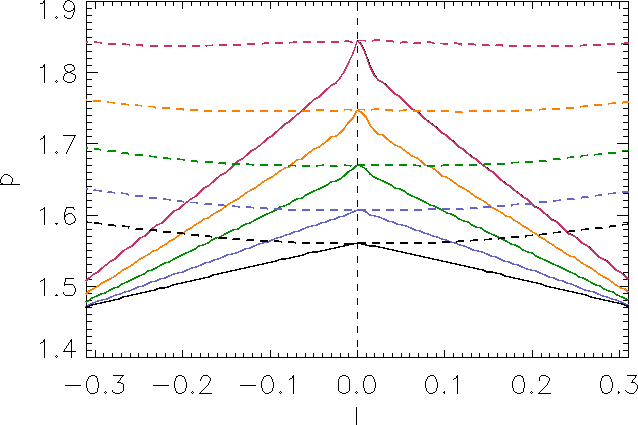}}
    \vspace{-0.325\textwidth}
\centerline{\Large \bf \hspace{0.03 \textwidth}  \color{black}{\small(a)}\hfill}
    \vspace{0.275\textwidth}
\centerline{\Large \bf \hspace{0.03 \textwidth}  \color{black}{\small(b)}\hfill}
  \vspace{0.01\textwidth}
\caption{Pressure (a) along the $z$-axis, normalised to the length of the respective separator, and (b) in a cut through the depth (solid) and across the width (dashed) of the current layer in the plane $z = 0.5$ for experiments with initial current $j_{sep} =$ 0.75 (black), 1.00 (blue), 1.25 (green), 1.5 (orange) and 1.75 (red).}
\label{fig:pressure}
    \end{figure}
    As already seen, in the final equilibrium state, cuts in planes perpendicular to the separator show that the separatrix surfaces collapse creating cusp-shaped regions about the separator, within which lie regions of enhanced pressure and outwith which the pressure falls off. The pressure difference (the pressure minus the initial pressure, $p - p_{0}$) in the $z = 0.5$ plane is shown in Fig.~\ref{fig:j1.5_pressure}a, for the experiment with initial current $j_{sep} = 1.5$. The deformation of the separatrix surfaces to produce cusps at the ends of the current layer is analogous to that seen in 2D when the separatrices of a 2D null collapse to form a current layer \citep[e.g.,][]{Klapper98,Craig05,FF11}. The resulting pressure enhancements, within the two cusp regions are also reminiscent of these 2D current layers. The cusp regions form due to the requirement that total pressure must balance across the current layer in an equilibrium state. From the lower-left and upper-right flux domains in Fig.~\ref{fig:j1.5_pressure}a, the magnetic field approaching the current layer tends to zero, but from the other two flux domains it tends to a finite value. For total pressure balance, the plasma pressure must be higher near the current layer in the first pair of domains than in the latter pair. The two flux domains with higher pressure form cusp regions as the magnetic field and pressure form a spiked wedge between the two other domains that lie almost parallel near the separator.

    Fig.~\ref{fig:j1.5_pressure}b shows the 3D extent of the regions of enhanced (yellow) pressure that occur inside the cusp regions about the separator and the pressure outside the cusps which falls off away from the separator (blue). In particular, it is clear that the four regions extend beyond the ends of the separator,  where one or other of the separatrix surfaces is bounded by a spine. The pressure difference weakens in these areas as you get further above or below the nulls off the ends of the separator, so it is possible that, if the domain was much longer, the pressure would reduce to uniform far away from the ends of the separator. 
    
     The resulting variation in plasma pressure in the final equilibrium obviously effects the plasma beta within the system. From Fig.~\ref{fig:j1.5_pressure}c, which displays contours of the plasma beta in the cut at $z=0.5$ for the equilibrium state, it is apparent that the enhanced regions of beta are confined to within the cusps close to the separator instead of being high anywhere within the vicinity of the separator (c.f. Fig.~\ref{fig:j1.5_beta}). Furthermore, the overall plasma beta in the system is slightly lower ($\beta = 6.9$) than it was initially.

There is enhanced plasma pressure along the length of the separator itself (Fig.~\ref{fig:pressure}a), producing a pressure gradient and, hence, a pressure force, as already discussed in Sect.~\ref{sec:tot}. To enable all the experiments to be compared, the lengths of the separators have all been normalised to one in the same way as they were for Fig.~\ref{fig:current}. The pressure enhancement along the separator is greatest in the experiment with the highest initial current and in all cases reaches its peak at about 57\%-58\% of the way along the length of the separator. Beyond the ends of the separator, along the $z$-axis, the plasma pressure becomes constant, but the plasma pressure along the $z$-axis above the upper null is slightly higher than it is below the lower null. Again, we suspect this is due to asymmetries in the initial field and plan to investigate this further in the future.

A cut through the depth of the current layer, in the plane $z=0.5$, reveals that the plasma pressure peaks at the separator, whilst in a cut across its width the pressure is almost constant (Fig.~\ref{fig:pressure}b). This behaviour agrees with that seen in the 2D cut of the pressure in the $z=0.5$ plane (Fig.~\ref{fig:j1.5_pressure}a) and indicates that in the immediate vicinity of the current layer there is a plasma-pressure gradient opposing the collapse of the current layer (also seen in Fig.~\ref{fig:j1.5_forces}b). The details of the small residual forces that remain in the equilibrium state of each experiment are discussed next.

\subsection{Forces through the depth and across the width of the current layer}\label{sec:forces}
\begin{figure}[!h] 
\centerline{\includegraphics[width=1.0\linewidth]{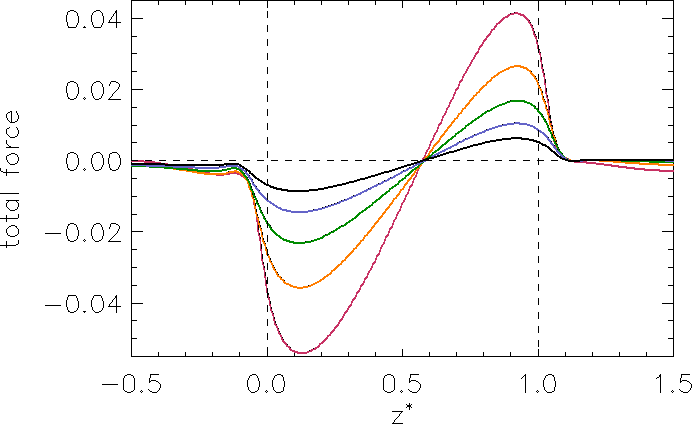}}
\centerline{\includegraphics[width=1.0\linewidth]{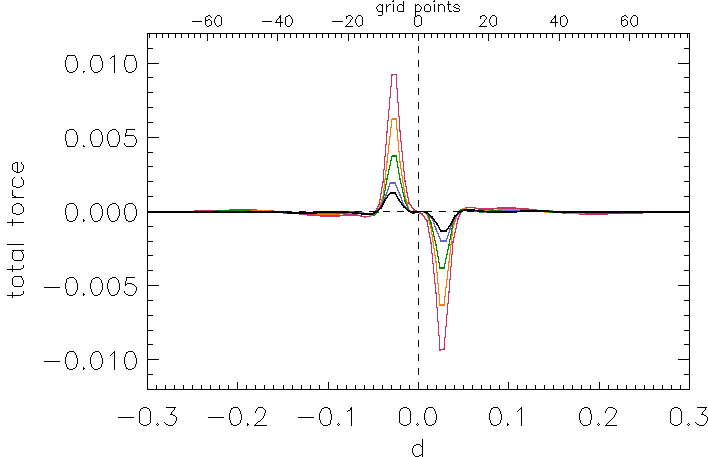}}
\centerline{\includegraphics[width=1.0\linewidth]{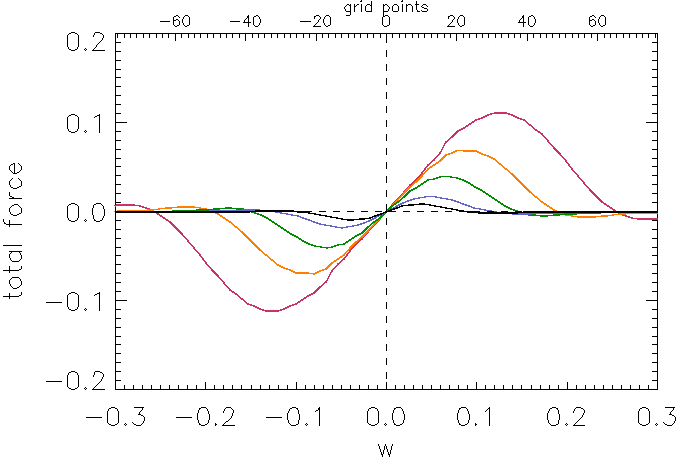}}
  \vspace{-0.675\textwidth}
\centerline{\Large \bf \hspace{0.03 \textwidth}  \color{black}{\small(a)}\hfill}
    \vspace{0.3\textwidth}
\centerline{\Large \bf \hspace{0.03 \textwidth}  \color{black}{\small(b)}\hfill}
    \vspace{0.3\textwidth}
\centerline{\Large \bf \hspace{0.03 \textwidth}  \color{black}{\small(c)}\hfill}
    \vspace{0.01\textwidth}
\caption{(a) The total force along the separator normalised to the length of the respective separator. Slices showing the total force  (b) through the depth and (c) across the width of the current layer in the plane $z = 0.5$. The colours represent the experiments with initial current $j_{sep} = 0.75$ (black), 1.00 (blue), 1.25 (green), 1.5 (orange) and 1.75 (red).}
\label{fig:allj_totforce}
    \end{figure}
    As already seen from Fig.~\ref{fig:j1.5_totalforce}, each experiment reaches a state in which all the forces balance everywhere within the domain, except within the current layer itself and along the separatrix surfaces.  Here, the residual forces are both small (in comparison to the initial forces) and highly localised. We call this the `equilibrium' state, although the field is actually only in a quasi-equilibrium. In the equilibrium state, the Lorentz force vanishes along the separator which means that the total force here is simply the pressure force (Fig.~\ref{fig:allj_totforce}a - here the length of each separator is normalised to one). The behaviour of the total force along the $z$-axis in the final equilibrium state is the same in each experiment: it acts outwards towards both nulls along the separator, from the same point just over half way along the separator where the plasma pressure reaches a maximum. However, the magnitude of this force increases with $j_{sep}$. 

The total force through the depth of the current layer acts inwards towards the separator such as to squeeze the current layer thinner (Fig.~\ref{fig:allj_totforce}b), whilst the total force across the width acts outwards away from the separator (Fig.~\ref{fig:allj_totforce}c). Naturally, in both figures the total force is seen to increase with increasing initial current $j_{sep}$. This behaviour of the total force perpendicular to the separator, which acts to perpetuate the collapse of the separator is the same as that seen in current layers formed from the collapse of a 2D null \citep[e.g.,][]{FF11}. 

\subsection{Growth rate of the current layer}\label{sec:growth}
These small, non-zero, and highly localised forces about the separator indicate that the current layer itself is not yet in equilibrium, even though the rest of the system is. Indeed, as already mentioned, it is possibly undergoing an infinite-time collapse, as is seen during the collapse of null points, in both 2D and 3D \citep{Klapper98,Pontin05,FF11,FF12,FF13}. Here, we investigate how the current grows within the current layer and how the initial current $j_{sep}$ affects this.

Fig.~\ref{fig:allj_modj} shows that, in each experiment, the maximum value of $|{\bf {j}}|$ in the separator slowly grows in time throughout the relaxation, following a time evolution of the form
\begin{equation}
|{\bf {j}}| = j_{sep}\Bigg(1 + a_{0}\frac{t}{t_{f}}\Bigg)^{a_{1}}.
\label{eq:growth}
\end{equation}
This form of growth is the same as that seen in the collapse of 2D and 3D nulls and is suggestive that there is an infinite-time singularity along the separator implying that the system is attempting to reach a true singularity which it cannot achieve in a finite time. Since we have followed the time evolution for one order of magnitude increase in time, however, we cannot be certain. The growth rate, $a_{1}$, is proportional to the initial uniform current $j_{sep}$, and in all cases considered here is less than 0.5. The same trend is found for the growth of the minimum value of $|{\bf {j}}|$ along the separator. In each experiment the maximum value of $|{\bf {j}}|$ occurs around $z = 0.4$ and the minimum values occur around the upper null.
\begin{figure}
\centering
\includegraphics[width=0.45\textwidth,clip]{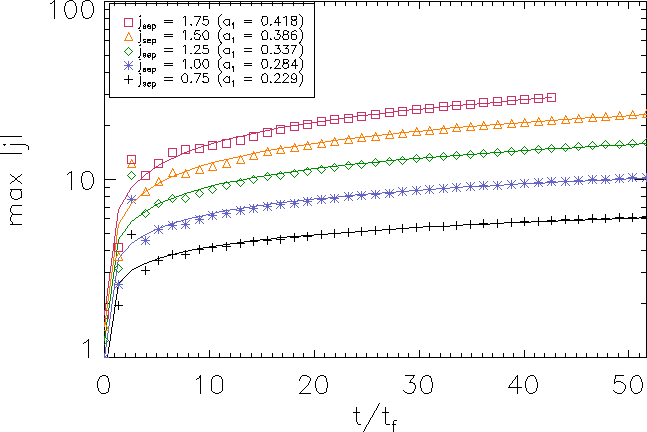}
\caption{Maximum values of $|{\bf {j}}|$ along the separator as a function of time with curves of Eq.~\ref{eq:growth} for experiments with initial current $j_{sep} =$ 0.75, 1.0, 1.25, 1.5 and 1.75.}\label{fig:allj_modj}
\end{figure}

\section{Conclusions}\label{sec:conc}
In this paper, we have performed the first non-resistive MHD relaxation of a single non-potential separator. 
Here, we have analysed the results from five experiments with varying initial uniform current in which an initially non-equilibrium magnetic field, containing two null points, their associated spines and separatrix surfaces and a separator connecting the two nulls, is allowed to evolve to an equilibrium state. These experiments determine where the current layer forms in a magnetic field containing various different topological features and what the characteristics of the current layer are.

In our experiments, the main current layers formed are centred on the separator. Separators are important topological features since, due to their position at the boundary of four topologically distinct flux domains, current builds up easily along them, as seen in our numerical experiments. Isosurfaces of the current reveal that the current layer is essentially a flat twisted band about the separator. However, lower isosurfaces of current reveal a more complex shape similar to that of a hyperbolic flux tube. At the ends of the separator, near the nulls, cross-sections perpendicular to the separator through the enhanced current regions are elongated ellipses that are aligned with the separatrix surface of the null nearest to the cross-sectional cut. In the middle of the separator the cross-sectional cuts have an X-type shape as weak wings of current are found extending along both separatrix surfaces. The separator current layers formed twist about the separator. Their degree of twist is dependent on the strength of the initial current. The current layer is twisted due to the fact that the initial uniform current is aligned with the separator causing the separatrix surfaces to twist about the separator.

The current accumulations along the separator are non-uniform, probably due to the initial asymmetries in the skeleton.  Also, the current profile along the $z$-axis possesses reverse currents outside the separator, as has been observed in some 2D current sheets. The dimensions (width, depth and length) of the current layer, as well as the amount of current in the current layer, are all found to depend on the initial current ${\bf {j}} = (0,0,j_{sep})$.

The final states of our experiments are all in equilibrium everywhere except near the separator and along the separatrix surfaces. In these highly localised regions, small residual forces remain causing the separator to slowly lengthen and widen throughout the relaxation, and also to continually flatten and strengthen in current. This slow, but continual evolution suggests the system is approaching an infinite-time singularity as is seen in the collapse of 2D and 3D nulls \citep[e.g.][]{FF11,FF12}. This would imply that a true equilibrium could not be achieved in a finite time. 

The plasma within the experiments starts off uniform, but, in the final equilibrium state, the separatrix surfaces about the separator have collapsed to form cusp regions in planes perpendicular to the separator: the plasma pressure builds up within the cusp regions and outwith them it falls off, as seen in the collapse of 2D nulls. Cusps of this nature are required to provide total pressure balance across the current layer.

The experiments considered here are the first such numerical models for separator current layers formed through non-resistive MHD relaxation. Here, we consider current layers arising from initial non-equilibrium magnetic fields with uniform current parallel to the separator and have observed that the current builds along the separator throughout the relaxation, as opposed to building at the null points. We would expect that having a smaller plasma beta would lead to higher currents building up at the separator current layer since the  pressure gradients that counteract the collapse of the separator would be weaker. We intend to investigate current layers formed in regions of low plasma beta in a follow-up paper. Furthermore, since other possible orientations for the current may effect the final equilibrium state, we will also study current layers created from different initial magnetic field configurations. 

Short length scales are necessary for 3D magnetic reconnection to take place, and so, current layers at separators, such as those formed here, are natural sites for 3D magnetic reconnection. In the future, we will use the equilibria formed here as initial states in order to study magnetic reconnection at separator current layers.

\bibliographystyle{aa}
\bibliography{julie}

\appendix
\section{Analytical magnetic field}\label{sec:app}
The initial analytical magnetic field was chosen as it represents a field with two 3D-null points whose separatrix surfaces intersect to form a single separator connecting the nulls. The field has constant current in the $z$ direction parallel to the separator. It was formed by starting with the lowest order (quadratic) magnetic field that represents two nulls joined by a separator. Such a field contains 27 unknown parameters since, for each component of the magnetic field ($B_{x}$, $B_{y}$, $B_{z}$), there are 9 terms ($x$, $y$, $z$, $xy$, $xz$, $yz$, $x^{2}$, $y^{2}$, $z^{2}$). Without loss of generality most of these terms can be eliminated by satisfying a series of conditions. The conditions that we impose on our field are as follows,
\begin{itemize}
\item $\nabla \cdot {\bf{B}} = 0$.
\item ${\bf{j}} = j_z\hat{{\textbf{z}}}$, so the current is constant and is directed along the separator.
\item ${\bf{B}} = {\bf{0}}$ only at $x = y = z = 0$ and at $x = y = 0, z = L$ to give two nulls a distance $L$ apart.
\item Only one separator exists and it lies along the $z$-axis.
\item The lower/upper null is positive/negative with a vertical separatrix surface and spine lying in the $z=0$/$z=L$ planes.
\end{itemize}
Satisfying these conditions allows the general field with 27 parameters to be reduced to a field of the form shown in Eq.~(\ref{eq:magfield}) with just five parameters (note, the magnetic field and length of the system have scaling factors $B_{0}$ and $L_{0}$ which are set equal to one here).
\begin{eqnarray}
\left. \begin{array}{rcl}
B_{x} &=& x+cxz+byz-\tfrac{1}{2} j_{sep}y, \\
B_{y} &=& (2a-c)yz-(1+La)y+bxz+\tfrac{1}{2} j_{sep}x, \\
B_{z} &=& a(Lz-z^2)+\tfrac{1}{2} cx^2+(a-\tfrac{1}{2} c)y^2+bxy.
\end{array}\right\}
\label{eq:magfield}
\end{eqnarray}
The length of the separator, $L$, is set to one in all experiments considered here. The four parameters $a$, $b$, $c$ and $j_{sep}$ have constraints on them in order to satisfy the conditions listed previously. The constraints are
\begin{itemize}
\item $a$ $\textgreater$ 0,
\item $b^{2}$ $\textgreater$ $c(2a - c)- \tfrac{(2a-acL-2c)^{2}}{j_{sep}^{2}-4-4aL}$,
\item $b^{2}$ $\textgreater$ $\tfrac{(1+c)(a-c-1)}{L^2} + \tfrac{j_{sep}^2}{4L^2}$,
\item $j_{sep}^{2}$ $\textless$ $4(1+aL)$.
\end{itemize}
Varying the parameters $a$ and $c$ modifies the geometry of the field lines in the separatrix surfaces of both nulls. Varying the parameter $b$ rotates the upper null's separatrix surface relative to the lower null's separatrix surface. Finally $j_{sep}$, the non-potential parameter, allows the separatrix surfaces of both nulls to curl around the separator.

\begin{acknowledgements}
JEHS would like to thank STFC for financial support during her Ph. D and CEP acknowledges support from the STFC consolidated grant. We would like to thank the referee for useful comments.
\end{acknowledgements}

\end{document}